\documentstyle[editedvolume,psfig]{crckapb}

\begin{opening}
\title{SCUBA'S FIRST-BORN: SMM\,J02399$-$0136}

\author{ROB IVISON}
\institute{Dept of Physics and Astronomy, University College London}
\author{IAN SMAIL}
\institute{Department of Physics, University of Durham}
\author{ANDREW BLAIN}
\institute{Cavendish Laboratory, Madingley Road, Cambridge}
\author{JEAN-PAUL KNEIB}
\institute{Observatoire de Toulouse, 14 avenue E.\ Belin,
           31400 Toulouse}
\author{DAVID FRAYER}
\institute{Astronomy Department, California Institute of Technology}

\end{opening}

\runningtitle{SMM\,J02399$-$0136}

\begin{document}

\section{Introduction}

We discuss observations of the submm-selected galaxy,
SMM\,J02399$-$0136, and what has been learnt about it during the year
following its discovery.  SMM\,J02399$-$0136 was the first distant
galaxy detected in submm surveys with SCUBA. Its association with a
massive, gas-rich starburst/AGN at $z=2.8$ has lead to suggestions
that the prevalence of AGN in the early Universe may be high (Ivison
et al.\ 1998) and that these AGN may account for a significant
fraction of the far-IR background.

\section{Discovery}

The discovery of SMM\,J02399$-$0136 (Ivison et al.\ 1998) came as a
surprise to all concerned, with the possible exception of Andrew Blain
who had been a long-time proponent of submm imaging of the distant
Universe using massive cluster lenses (Blain 1997). The discovery
images were obtained with SCUBA during uncharacteristically good
weather in the summer of 1997 by Smail, Ivison \& Blain (1997). As
often seems to happen, SMM\,J02399$-$0136 was seen in the first map,
behind the $z=0.37$ massive cluster, Abell~370. The area covered
during that first night has since increased by two orders of
magnitude, with the completion of the SCUBA Lens Survey (Smail et al.\
1998; Blain et al.\ 1999; Smail et al., these proceedings) and the
commencement of several large, conventional blank-field surveys (e.g.\
Eales et al.\ 1999) but SMM\,J02399$-$0136 remains the brightest
submm-selected galaxy, by virtue of its amplification by the
foreground cluster (a factor $2.4 \pm 0.3$).  This amplification aids
us in the follow-up of SMM\,J02399$-$0136 at all wavelengths, and when
combined with the lavish archival datasets available for this field,
has allowed a detailed view of the nature of this source to be
achieved relatively quickly.

%
%
\begin{figure}
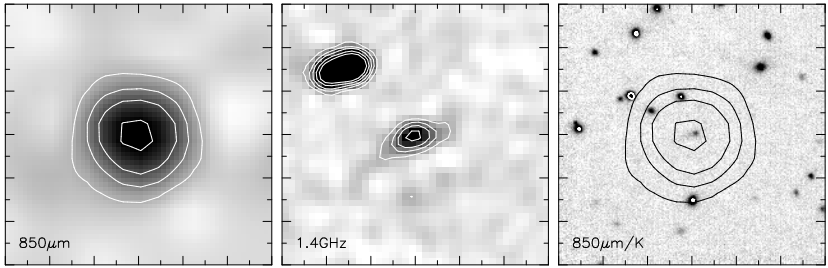

\centerline{\psfig{file=ivison_fig1a.ps,angle=270,width=4.3in}}
\vspace*{-0.15cm}
\caption{{\em Top:} 1$'$-square maps of SMM\,J02399$-$0136: 850-$\mu$m
SCUBA image, 1.4-GHz VLA image, and the 850-$\mu$m map overlaid on a
UKIRT $K$-band image. {\em Bottom:} 20$''$ square $BV\!RI\!K$ images.
L1 appears to be very compact, even at the very fine 0.4--0.6$''$
resolution of the CFHT $BRI$ exposures. L2 may be either a companion
or the remnant of a tidal interaction. L2 is most clearly visible in
$B$ and $K$, which may reflect a high emission-line contribution
(Ly$\alpha$; [O{\sc iii}]\,4959, 5007 with Balmer\,$\alpha$).}
\vspace*{0.2cm}
\centerline{\psfig{file=ivison_fig1b.ps,angle=270,width=4.7in}}
\end{figure}

\section{New and archival data}

Fortuitously, a deep (10\,$\mu$Jy\,beam$^{-1}$) 1.4-GHz map of A\,370
obtained some years ago by Frazer Owen and K.\,S.\ Dwarakanath
revealed a weak, extended radio counterpart within the error box of
the submm position of SMM\,J02399$-$0136 (Fig.~1).  A pair of optical
counterparts, resolved in archival CFHT images (Kneib et al.\ 1994),
are within $1''$ of the radio source. L1, the compact component, is
marginally resolved with an intrinsic FWHM of 0.3$''$. L2 has a more
complex morphology than L1, showing a ridge of emission to the north
and a diffuse region extending south and west towards L1.  L1 and L2
are separated by $\sim3''$ ($\sim 9$\,kpc after correcting for
tangential amplification).

The swift provision of near- and mid-IR images from UKIRT and {\em
ISO} by Tim Naylor and Leo Metcalfe showed that at least one of the
two counterparts possessed a spectral energy distribution (SED) whose
broad features were consistent with those expected for a submm-bright
galaxy.

%
%
\begin{figure}
\centerline{\psfig{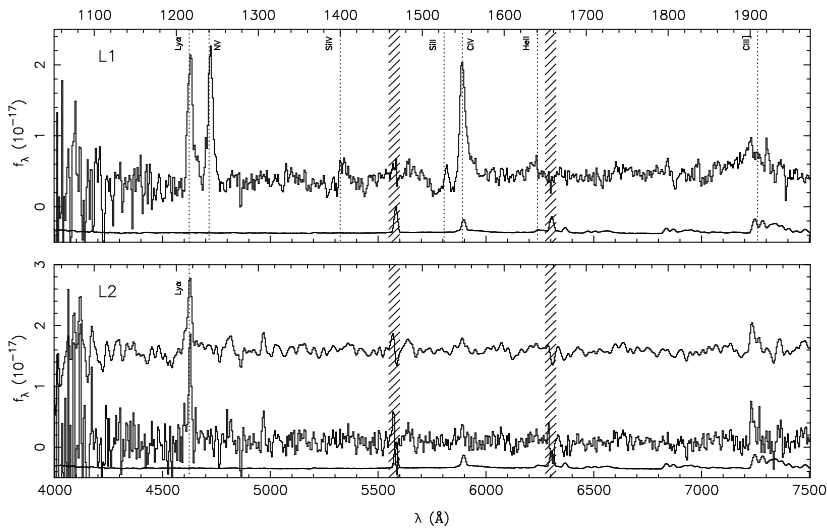}}
\vspace*{-0.3cm}
\caption{Optical spectra of L1 and L2 (Ivison et al.\ 1998). L1 shows
a number of narrow emission lines at $z=2.803$ superimposed on a blue
continuum, with hints of broad absorption features on the blue wings
of some lines. The substantially fainter L2 shows only Ly$\alpha$ and
Si\,{\sc ii}/O\,{\sc i} at a similar redshift to that of L1. The lower
spectrum in each panel is an arbitrarily scaled sky spectrum; the
hatched regions are strongly affected.}
\end{figure}

%
%
\begin{figure}
\centerline{\hspace*{0.7cm} \psfig{file=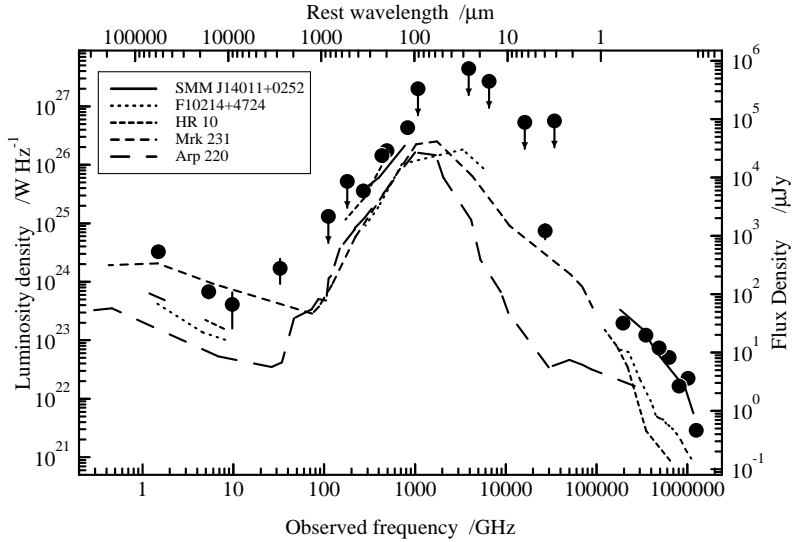,width=3.7in}}
\vspace*{-0.2cm}
\caption{SED of SMM\,J02399$-$0136 (filled circles). The right-hand
scale gives the flux densities for this galaxy. For comparison, we
have plotted the SEDs of the only other reliably identified submm
galaxy, SMM\,J14011+0252 at $z=2.6$ (Ivison et al.\ 1999), several
{\em IRAS} galaxies, and the $z=4.25$ radio galaxy, 8C\,1435+635, with
units of luminosity density on the left-hand scale. Amplification
corrections have been applied. For SMM\,J02399$-$0136, data were taken
from Ivison et al.\ (1998), A.\ Cooray, K.\ Dwarakanath, L.\ Metcalfe
and F.\ Owen (priv.\ comm.) and Frayer et al.\ (1998).}
\end{figure}

Since the optical counterparts were relatively bright ($I_{\rm total}
\sim 20.5$, $22.7$), we added a slit to a mask being using for
multi-object spectroscopy of A\,370 with the CFHT and obtained
high-quality optical spectra (Fig.~2). These clearly show that both
counterparts are at the same redshift, $z=2.803 \pm 0.003$.  Both have
faint continua with narrow lines in emission: L1 shows strong, narrow
Ly\,$\alpha$, N\,{\sc v} and C\,{\sc iv}, hints of weak Si\,{\sc ii},
Si\,{\sc iv}, He\,{\sc ii} and possibly a broad C\,{\sc iii}] line; L2
shows only weak, narrow Ly\,$\alpha$ and Si\,{\sc ii}/O\,{\sc i}, with
the Ly\,$\alpha$ emission extending over at least 8$''$.

The 1.4-GHz radio emission covers $7.9'' \times 2.2''$, with a
position angle (PA) of 71$^{\circ}$, a maximum surface brightness of
221\,$\mu$Jy\,beam$^{-1}$ and an integrated flux density of $526 \pm
50$\,$\mu$Jy. This is below the detection thresholds of most radio
surveys, even after lens amplification. The rest-frame
far-IR-to-5\,GHz flux ratio is similar to that seen in nearby
starbursts (Condon et al.\ 1991), which could be taken as evidence
that a starburst is the dominant contributor to the far-IR luminosity;
however, a recent 5-GHz map shows a PA closer to the optical/IR
morphology, which suggests that the 1.4-GHz emission may be from the
AGN.

Near-IR spectra of [O\,{\sc iii}] and Balmer\,$\alpha$ were also
obtained. Observations of both lines were extremely challenging, as
the atmosphere at 1.9 and 2.5\,$\mu$m is a better door than a window.
Only modest detections were obtained, suggesting narrow cores to the
lines; however, there is little hope of detecting broad components (if
present).

The overall SED of SMM\,J02399$-$0136 is shown in Fig.~3.  L1 and L2
both have smooth, steep UV--optical--mid-IR continua.  Between 120 and
350\,$\mu$m (rest frame), the SED has the characteristic spectral
index $\alpha \simeq +3$ of optically-thin emission from dust
grains. The far-IR luminosity (20--1000\,$\mu$m), $L_{\rm FIR} \sim
10^{13}$\,L$_{\odot}$ (after correcting for lensing).  The dust mass
is $5\times 10^8$\,M$_{\odot}$ for $T_{\rm d} = 50$\,{\sc k}. If the
dust is heated primarily by OB-type stars then $L_{\rm FIR}$
corresponds to an SFR ($>10$\,M$_\odot$ stars) of
$\sim2000$\,M$_{\odot}$\,yr$^{-1}$ ($\sim
6000$\,M$_{\odot}$\,yr$^{-1}$ if the IMF extends down to much lower
masses). Similarly high estimates of the SFR are given by the
H$\alpha$ luminosity (2000--20000\,M$_{\odot}$\,yr$^{-1}$) and the
radio luminosity, which predicts a supernova rate of
80--400\,yr$^{-1}$.  By any standards this would be a spectacular
starburst.

%
%
\begin{figure}
\centerline{\psfig{file=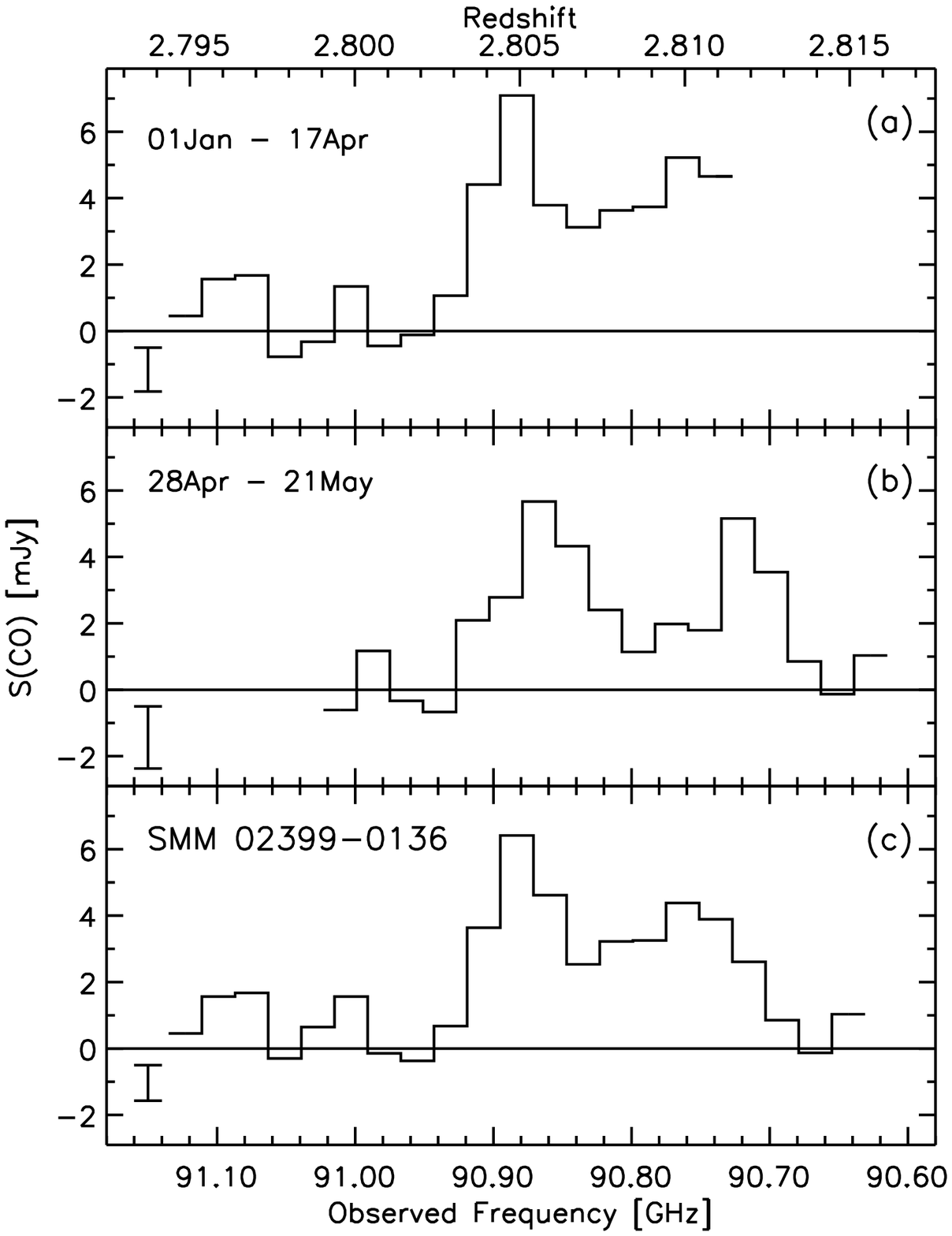,width=2.2in}\hspace*{1.0cm}
\vspace*{-0.2cm}
\psfig{file=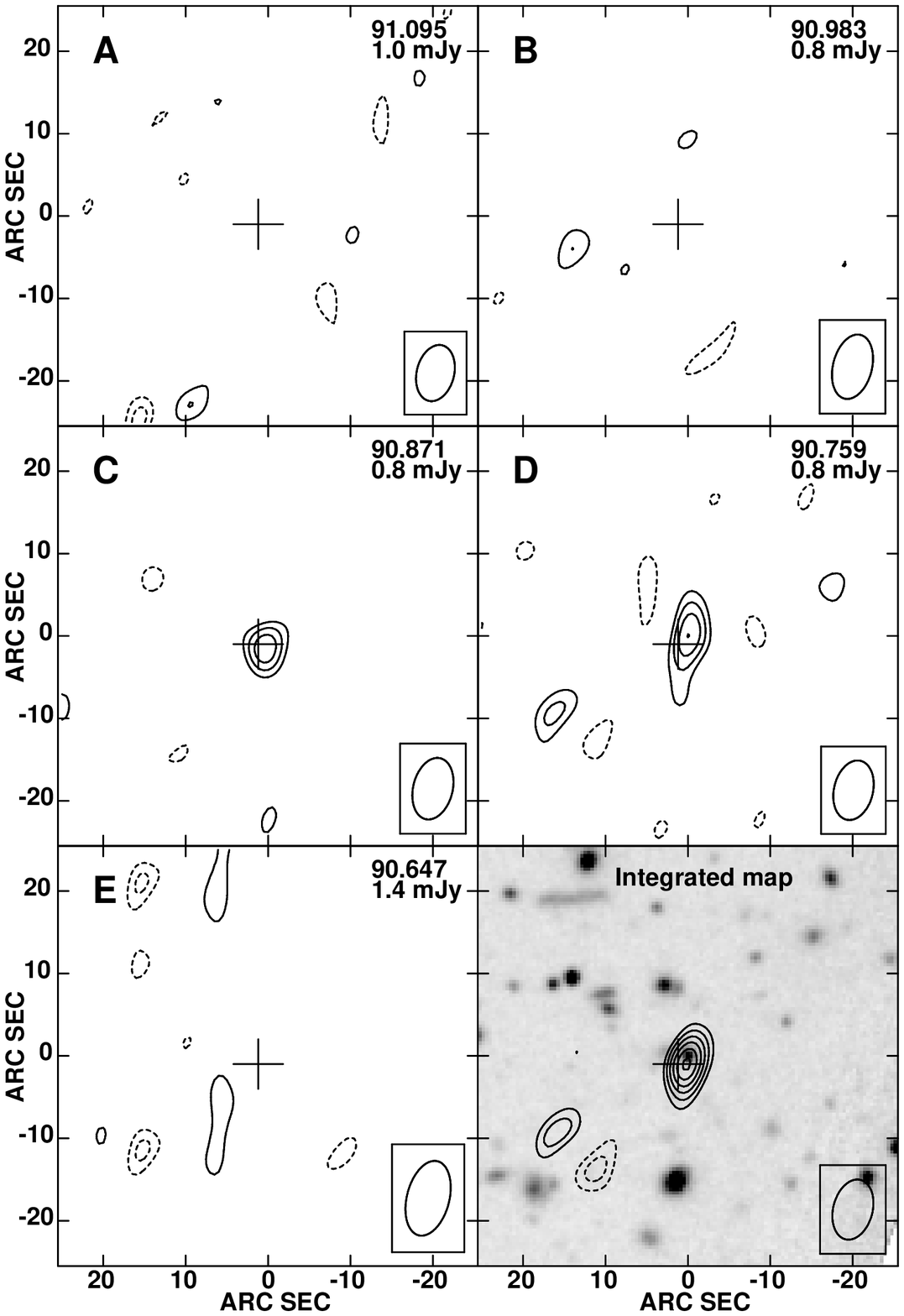,width=2.0in}}
\caption{{\em Left:} Spectra of SMM\,J02399$-$0136 observed with the
OVRO Millimeter Array (Frayer et al.\ 1998).  Panels (a) \& (b) show
subsets of the data, while panel (c) displays the cumulative
spectrum. {\em Right:} Panels A--E show channel maps averaged over
112\,MHz.  In the lower right panel is the integrated CO map overlaid
on a $B$-band image.}
\end{figure}

The most recent observational success was a search for molecular gas
in the system (Frayer et al.\ 1998). The search began at the optical
redshift, using the Owens Valley Millimeter Array. After 38\,hr of
integration time, a weak signal with coherent phases was found at the
reddest velocities. A further 16\,hr was spent at a lower frequency to
obtain the complete line profile shown in Fig.~4. The CO emission is
unresolved ($<5''$) and positionally coincident with L1. It is
redshifted by 400\,km\,s$^{-1}$ with respect to the optical lines,
with $z_{\rm CO} = 2.808$. The line is broad ($710\pm
80$\,km\,s$^{-1}$), with an apparent double-peaked profile.

The high molecular gas mass implied by the data
($\sim10^{11}$\,M$_{\odot}$) lends weight to arguments that a
significant fraction of the immense far-IR luminosity is due to star
formation. Such a mass is not unique for high-redshift systems but it
is several times higher than the most luminous low-redshift {\em IRAS}
galaxies, implying that SMM\,J02399$-$0136 will evolve into a massive
galaxy. The large gas mass, compared to the dynamical mass, suggests
that the gas is a dynamically important component of this galaxy and
points to its relative youth. On the other hand, the gas-to-dust ratio
(400 for $T_{\rm dust} = 50$\,{\sc k}) is similar to that found for
other high-redshift CO sources, suggesting that like many other
high-redshift massive galaxies, SMM\,J02399$-$0136 is already
chemically evolved.

SMM\,J02399$-$0136 is one of two galaxies from the SCUBA Lens Survey
to be detected in CO to date, the other being SMM\,J14011+0252 at
$z=2.6$ (Ivison et al.\ 1999; Frayer et al.\ 1999).  These are the
first two members of the submm field population to be investigated in
detail. Their optical emission-line characteristics are radically
different, with one showing strong AGN signatures, the other an
apparently pure starburst spectrum; however, both are found to be
associated with gas-rich, massive galaxies, which supports the idea
that a significant proportion of the submm galaxy population is made
up of proto-ellipticals.

In summary, SMM\,J02399$-$0136 shows clear signs of the presence of an
AGN, both in its optical emission-line properties and its radio
morphology.  However, there are also indications of an on-going
starburst: extended optical emission, narrow and strong H$\alpha$
emission, a large mass of dust and a dynamically significant gas
reservoir.  If asked the question: ``Is SMM\,J02399$-$0136 an AGN or a
starburst?'', we'd probably have to answer: ``Both''.  Critical tests
of the relative luminosity of the AGN and the starburst include the
identification in polarized light of hidden broad-line components to
the rest-frame UV/optical emission lines, a search with {\it AXAF} for
hard X-ray emission (which should escape an obscured active nucleus),
and high-resolution 1.4-GHz images to look at the radio emission
characteristics in more detail.

We must wait and see whether a significant fraction of submm-selected
galaxies resemble SMM\,J02399$-$0136. A large AGN contribution to the
far-IR background would certainly resolve potential problems
concerning over-production of metals, though there are other solutions
--- modifying the IMF, for example (Blain et al.\ 1999). A decisive
test of the contribution of AGN-powered emission to the extragalactic
background awaits the detailed study of a representative sample of the
submm-selected galaxies that dominate the submm background
emission. The faintness of these sources in the optical/near-IR and
millimetre wavebands compared to the sensitivities of current
instrumentation means that the advantages of using lens amplification
will probably remain clear for these important studies.

\section*{Acknowledgements}

We acknowledge support from PPARC and the Royal Society and thank
Jacqueline Davidson, Tom Jones and Plaid Cymru for inspiration.


\begin{thebibliography}{}
\bibitem[]{} Blain, A.W., 1997, MNRAS, 290, 553{}{}
\bibitem[]{} Blain, A.W.\ et al., 1999, MNRAS, 302, 632{}{}
\bibitem[]{} Condon, J.J., Frayer, D.T., Broderick, J.J., 1991, AJ, 101, 362{}{}
\bibitem[]{} Eales, S.A.\ et al., 1999, ApJ, in press (astro-ph/9808040){}{}
\bibitem[]{} Frayer, D.T.\ et al., 1998, ApJ, 506, L7{}{}
\bibitem[]{} Frayer, D.T.\ et al., 1999, ApJ, in press (astro-ph/9901311){}{}
\bibitem[]{} Ivison, R.J.\ et al., 1998, MNRAS, 298, 583{}{}
\bibitem[]{} Ivison, R.J.\ et al., 1999, MNRAS, submitted{}{}
\bibitem[]{} Kneib, J.-P.\ et al., 1994, A\&A, 286, 701{}{}
\bibitem[]{} Smail, I., Ivison, R.J., Blain, A.W., 1997, ApJ, 490, L5{}{}
\bibitem[]{} Smail, I.\ et al., 1998, ApJ, 507, L21{}{}
\end{thebibliography}
\end{document}